\documentclass[a4paper,11pt]{article}
\pdfoutput=1
\usepackage{graphicx}
\usepackage{fullpage}
\usepackage{jheppub} 
\usepackage{bbm}
\usepackage{amsfonts}
\usepackage{slashed}
\usepackage{subfigure}
\usepackage{array}
\usepackage{verbatim}
\usepackage{booktabs}
\usepackage{color, soul}
\usepackage{mathrsfs}
\usepackage{epsfig}
\usepackage{graphicx}
\usepackage{dcolumn}
\usepackage{bm}
\usepackage{amsmath}
\usepackage{amssymb}
\usepackage{multirow}

\setstcolor{red}

\usepackage{makecell}

\title{\sffamily Scalar dark matter interpretation of the DAMPE data with U(1) gauge interactions}

\author[a,b]{Junjie Cao}
\author[c]{,Lei Feng}
\author[a]{,Xiaofei Guo}
\author[a]{,Liangliang Shang}
\author[a,d]{,Fei Wang}
\author[e]{,Peiwen Wu}

\affiliation[a]{College of Physics and Materials Science, Henan Normal University, Xinxiang 453007, China}
\affiliation[b]{Center for High Energy Physics, Peking University, Beijing 100871, China}
\affiliation[c]{Key Laboratory of Dark Matter and Space Astronomy, Purple Mountain Observatory, Chinese Academy of Sciences, Nanjing 210008, China}
\affiliation[d]{School of Physics, Zhengzhou University, 450000, ZhengZhou, P.R.China}
\affiliation[e]{School of Physics, KIAS, 85 Hoegiro, Seoul 02455, Republic of  Korea}

\emailAdd{junjiec@itp.ac.cn}
\emailAdd{fenglei@pmo.ac.cn}
\emailAdd{guoxf@gs.zzu.edu.cn}
\emailAdd{shlwell1988@foxmail.com}
\emailAdd{feiwang@zzu.edu.cn}
\emailAdd{pwwu@kias.re.kr}

\abstract{Recently, DAMPE experiment released the new measurement of the total cosmic $e^+e^-$ flux between 25 GeV and 4.6 TeV which indicates a spectral softening at around 0.9 TeV and a tentative peak at around 1.4 TeV. We utilize the scalar dark matter (DM) annihilation scenario to explain the DAMPE peak by extending $G_{SM}\equiv SU(3)_C \times SU(2)_L \times U(1)_Y$ with additional $U(1)$ gauge symmetries while keeping anomaly free to generate $\chi \chi \to Z^\prime Z^\prime \to \ell\bar{\ell}\ell^\prime\overline{\ell^\prime}$, where $\chi, Z^\prime, \ell^{(^\prime)}$ denote the scalar DM, the new gauge boson and $\ell^{(^\prime)}=e,\mu,\tau$, respectively, with $m_\chi \sim m_{Z^\prime} \sim 2 \times 1.5$ (TeV). We first illustrate that the minimal framework $G_{SM} \times U(1)_{Y^\prime}$ with the above mass choices can explain the DAMPE excess but has been excluded by LHC constraints from the $Z^\prime$ searches. Then we study a non-minimal framework $G_{SM} \times U(1)_{Y^\prime} \times U(1)_{Y^{\prime \prime}}$ in which $U(1)_{Y^{\prime \prime}}$ mixes with $U(1)_{Y^\prime}$. We show that such a framework can interpret the DAMPE data while passing other constraints including the DM relic abundance, DM direct detection and collider bounds. We also investigate the predicted $e^+e^-$ spectrum in this framework and find that the mass splitting $\Delta m = m_\chi - m_{Z'}$ should be less than about 17 GeV to produce the peak-like structure.}

\begin{document}
\maketitle \indent
\newpage

\section{\label{introduction}Introduction}

Recently, the DArk Matter Particle Explorer (DAMPE) experiment released the new measurement of the total cosmic $e^+e^-$ flux between 25 GeV and 4.6 TeV which indicates a spectral softening at around 0.9 TeV and a tentative peak at around 1.4 TeV \cite{Collaboration2017,TheDAMPE:2017dtc}. The spectral softening, as was pointed out in \cite{Yuan2017}, may be due to the breakdown of the conventional assumption of continuous source distribution or the maximum acceleration limits of electron sources, while the peak can be explained by dark matter (DM) annihilation in a nearby clump halo into exclusive $e^+e^-$
or equal $e,\mu,\tau$ branching fractions (Brs) with $\langle \sigma v \rangle \sim 3 \times 10^{-26} \,{\rm cm^3/s}$. The best
fit values for the DM particle mass, the DM clump mass and the annihilation luminosity $\mathcal{L}=\int \rho^2 dV$ are
around 1.5 TeV, $10^{7-8} \, M_{\rm sun}$ and $10^{64-66}\, {\rm GeV^2\,cm^{-3}}$ respectively, depending on the halo distance from earth.

Interestingly, the excesses in cosmic positron flux have been previously announced in AMS-02 \cite{Aguilar:2013qda,Accardo:2014lma}, PAMELA \cite{Adriani:2008zr,Adriani:2010ib} and Fermi \cite{FermiLAT:2011ab}. To interpret the positron anomaly both astrophysical (for a review see e.g. \cite{Serpico:2011wg}) and DM origins (for a review see e.g. \cite{Cirelli:2012tf}) have be proposed. Since no confirmed source has been established, in this paper we assume DM democratic annihilation into $e,\mu,\tau$ to explain the DAMPE data. This interpretation requires
the following four conditions:
\begin{itemize}
\label{list-conditions-1}
\item \textbf{I-ID}:  \textit{large} $\langle \sigma v \rangle_0 \gtrsim 1 \times 10^{-26} \,{\rm cm^3/s}$ with $v\sim 10^{-3}\,c$ in today's Universe, meanwhile coinciding with other DM indirect search constraints.
\item \textbf{II-RD}: \textit{large} $\langle \sigma v \rangle_{FO} \sim 1\times10^{-26} \, {\rm cm^3/s}$ with $v \sim 0.1 \,c$ in early freeze out.
\item \textbf{III-DD}:  \textit{small} DM-nucleon scattering $\sigma^{SI}_{DM-p}$ to pass DM direct detection bounds.
\item \textbf{IV-Collider}: \textit{small} signals to pass relevant collider constraints.
\end{itemize}

If the DM particle is a singlet under the Standard Model (SM) gauge group $G_{SM}\equiv SU(3)_C \times SU(2)_L \times U(1)_Y$, a mediator sector is needed to connect the DM and the SM. To satisfy $\langle \sigma v \rangle_0 \sim \langle \sigma v \rangle_{FO} \sim 1\times10^{-26} \, {\rm cm^3/s}$, the $s$-wave in $\langle \sigma v \rangle \simeq a + b \,v^2$ must be dominant to avoid $p$-wave suppression. One must also avoid chiral suppression given the light lepton masses (see e.g. \cite{Chang:2013oia,Berlin:2014tja}).
One solution is to make DM annihilate dominantly into bosons in the first step, followed by boson decays into leptons. If DM annihilates into scalar
pair, the new scalars usually mix with the SM Higgs which makes it difficult to generate nearly equal $e,\mu,\tau$ Brs. However, DM annihilation into new gauge vector boson pair $Z^\prime Z^\prime$ can easily produce the required decay Brs while being anomaly free. Moreover, scalar DM can benefit from DM-DM-$Z^\prime$-$Z^\prime$ contact interaction which is free of internal propagator suppression.

\begin{figure}[t]
\begin{center}
\includegraphics[width=5cm,height=5.1cm]{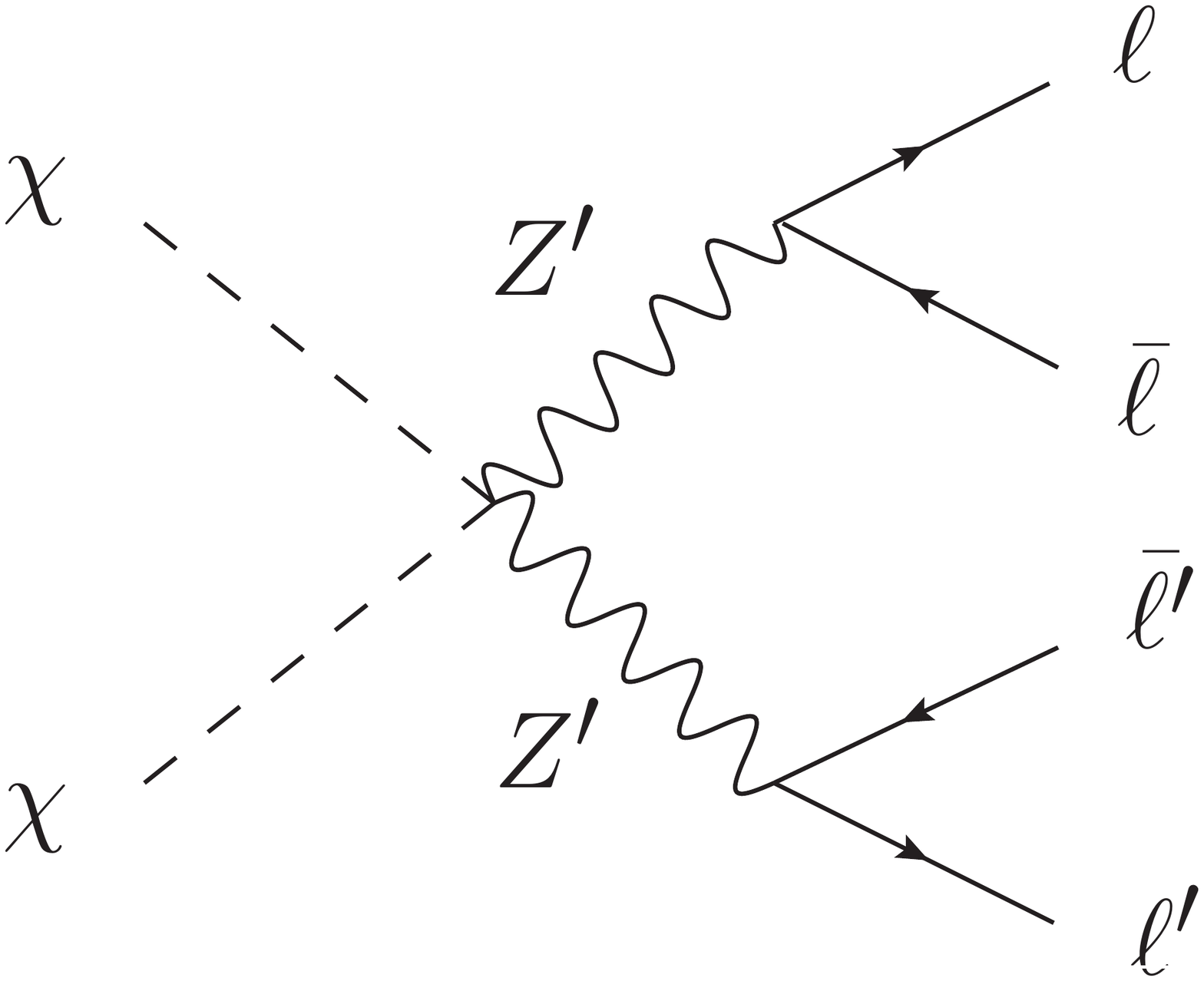}
\includegraphics[width=5cm,height=5.1cm]{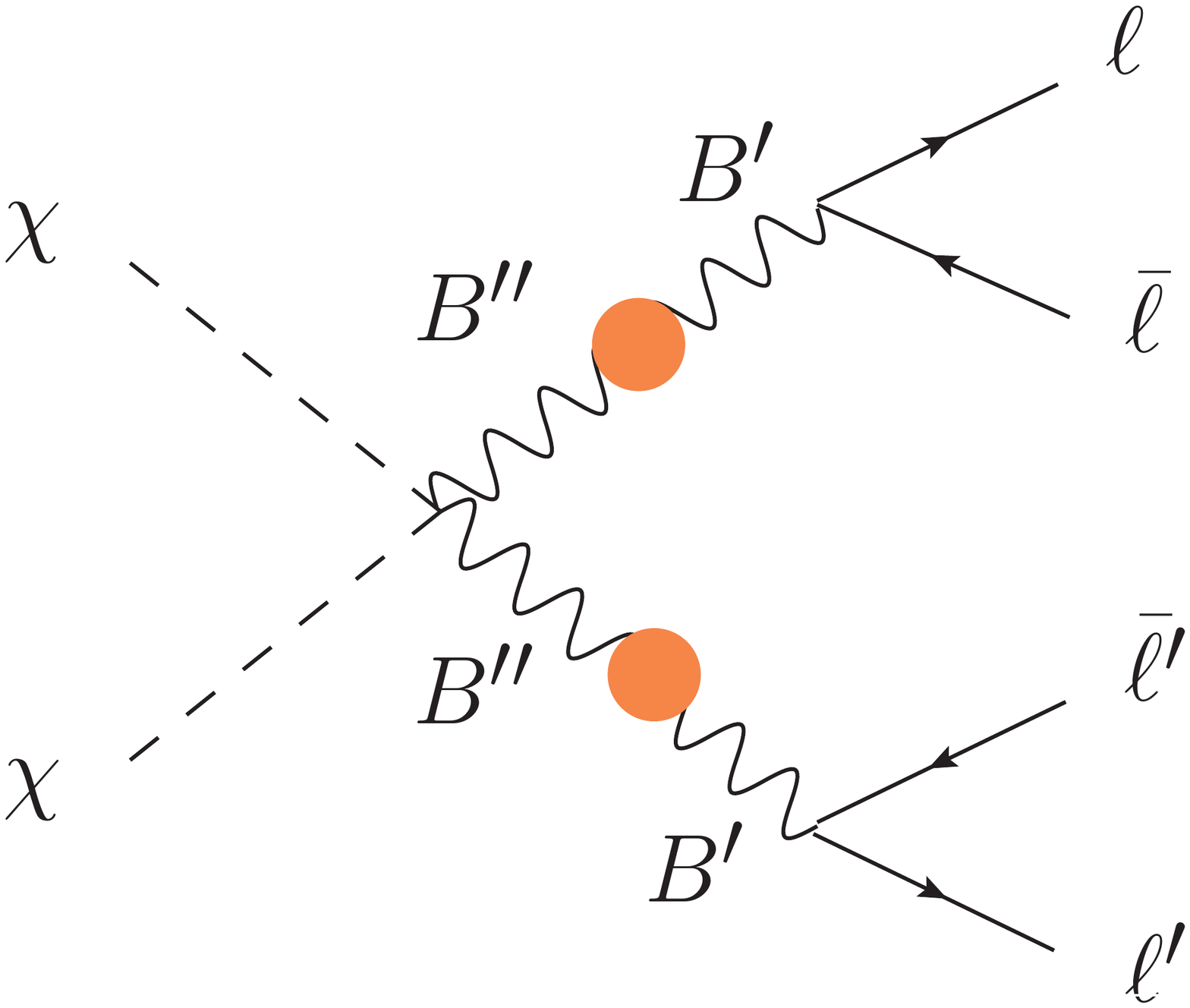}
\vspace{-1.0cm}
\caption{DM annihilation channels to interpret the DAMPE peak, where $\chi, Z^\prime,B^\prime,B^{\prime\prime}, \ell^{(^\prime)}$ denote the scalar DM, the new gauge bosons and $\ell^{(^\prime)}=e,\mu,\tau$, respectively. {\textbf{Left}}: model $G_{SM} \times U(1)_{Y^\prime}$; {\textbf{Right}}: model $G_{SM} \times U(1)_{Y^\prime} \times U(1)_{Y^{\prime \prime}}$ with dot indicating the mixing between $U(1)_{Y^{\prime \prime}}$ and $U(1)_{Y^\prime}$.}
\label{fig-omg}
\end{center}
\end{figure}

In this work we consider anomaly free theories and extend $G_{SM}$ with one or more additional $U(1)$ gauge symmetries to generate $\chi \chi \to Z^\prime Z^\prime \to \ell\bar{\ell}\ell^\prime\overline{\ell^\prime}$, where $\chi, Z^\prime, \ell^{(^\prime)}$ denote the scalar DM, the new gauge boson and $\ell^{(^\prime)}=e,\mu,\tau$, respectively, with $m_\chi \sim m_{Z^\prime} \sim 2 \times 1.5$ (TeV). We first illustrate that the minimal version $G_{SM} \times U(1)_{Y^\prime}$ with the above mass choices can explain the DAMPE excess but has been excluded by \textbf{IV-Collider}, i.e. LHC constraints from the $Z^\prime$ searches. Then we consider $G_{SM} \times U(1)_{Y^\prime} \times U(1)_{Y^{\prime \prime}}$ in which $U(1)_{Y^{\prime \prime}}$ mixes with $U(1)_{Y^\prime}$ to interpret the DAMPE data while passing all four conditions \textbf{I-ID, II-RD, III-DD, IV-Collider}. Fig.\ref{fig-omg} shows the DM annihilation channels in two models to interpret the DAMPE peak. We note that some early studies on the DAMPE peak in terms of fermionic DM which annihilates directly into leptons \cite{Yuan2017,Fang2017,Duan2017,Gu2017,Fan2017,Athron:2017drj} did not include the anomaly free condition on the theory, which are different from our assumption.

The paper is organized as follows. In Section \ref{Section-Model-1} we show the failure of the minimal extension $G_{SM} \times U(1)_{Y^\prime}$ when confronting the LHC constraints. In Section \ref{Section-Model-2} we discuss the framework with the symmetry $G_{SM} \times U(1)_{Y^\prime} \times U(1)_{Y^{\prime \prime}}$ which is successful in interpreting DAMPE data while satisfying other constraints. In Section \ref{Spectrum}, we investigate the $e^+e^-$ spectrum predicted in our scenario and study the upper bound
on the mass splitting between the DM and the gauge boson mediator if a peak-like structure is required. Finally, we
draw our conclusion in Section \ref{Section-Conclusion}.

\section{\label{Section-Model-1}$G_{SM} \times U(1)_{Y^\prime}$ framework}

In this section, we extend the $G_{SM}$ by one gauged $U(1)$ family symmetry to interpret the DAMPE peak.
Although there are many possible ways for the charge assignments to guarantee anomaly cancelation, we adopt the
minimal setting to predict equal $e,\mu,\tau$ coupling strengths for the new gauge boson, which was firstly proposed in \cite{Lee:2010hf}.
The particle content of the model with chiral anomaly cancelation is shown in Table.\ref{table-model-1},
where we introduce 3 generations of right handed (RH) neutrinos $\nu_{R,\{1,2,3\}}$ and two complex
scalars $\phi_s, \phi_\chi$. We impose an odd $Z_2$ parity for $\phi_\chi$ to generate a stable DM particle,
and require $\phi_s$ to develop a vacuum expectation value (vev) $v_s$ to generate $m_{Z^\prime}$. Since the neutrino
physics of the model has been intensively studied in \cite{Lee:2010hf}, we only concentrate on the DM physics to explain
the DAMPE peak in our discussion.

\begin{table}
\caption{Particle contents and their charge assignments in $G_{SM} \times U(1)_{Y^\prime}$ model.}
\begin{center}
\begin{tabular}{|c|c|c|c|c|c|c|}
\hline
Name 				& Spin	&  Gen. & $SU(3)_C$ & $SU(2)_L$ & $U(1)_Y$  & $U(1)_{Y^\prime}$ \\
\hline \hline
$H$						& 0 		& 1 	& {\bf 1} 					& {\bf 2} 		& -$\frac{1}{2}$ 	&  0 \\
\hline
$Q$						& 1/2 		& 3 	& {\bf 3} 					& {\bf 2} 		&  $\frac{1}{6}$ 	&  $\frac{1}{3}$ \\
$d_R^*$ 				& 1/2 		& 3 	& {\bf $\bar{\bf 3}$}	& {\bf 1} 		&  $\frac{1}{3}$ 	& -$\frac{1}{3}$ \\
$u_R^*$					& 1/2 		& 3 	& {\bf $\bar{\bf 3}$}	& {\bf 1} 		& -$\frac{2}{3}$	& -$\frac{1}{3}$ \\
\hline
$L_1$					& 1/2 		& 1 	& {\bf 1}					& {\bf 2} 		& -$\frac{1}{2}$ 	&  3 \\
$L_{\{2,3\}}$			& 1/2 		& 2 	& {\bf 1} 					& {\bf 2} 		& -$\frac{1}{2}$ 	& -3 \\
$\ell^*_{R,1}$			& 1/2 		& 1 	& {\bf 1} 					& {\bf 1} 		& 1 				& -3 \\
$\ell^*_{R,\{2,3\}}$	& 1/2 		& 2 	& {\bf 1}					& {\bf 1} 		& 1 				& 3 \\
\hline \hline	
$\nu^*_{R,1}$			& 1/2 		& 1	& {\bf 1}					& {\bf 1} 		& 0 				& -3 \\
$\nu^*_{R,\{2,3\}}$	& 1/2 		& 2 	& {\bf 1} 					& {\bf 1} 		& 0 				&  3\\
$\phi_s$ 				& 0		& 1 	& {\bf 1} 					& {\bf 1} 		& 0 				&  6 \\
$\phi_\chi$				& 0		& 1 	& {\bf 1} 					& {\bf 1} 		& 0 				&  6 \\
\hline
\end{tabular}
\end{center}
\label{table-model-1}
\end{table}

The most relevant Lagrangian in our mechanism includes
\begin{eqnarray}
\label{eqn-L-model-1}
{\mathcal L} \supset && | D^\prime_{\mu} \phi_\chi |^2 + | D^\prime_{\mu} \phi_s |^2 - V(H,\, \phi_\chi,\, \phi_s) \\ \nonumber
&& - \frac{1}{4} |F^\prime_{\mu\nu}|^2  + g_{Y^\prime} Z^\prime_{\mu} (Y^\prime_{f_R} \overline{f_R} \gamma^\mu f_R  + Y^\prime_{\nu_R} \overline{\nu_{R,i}} \gamma^\mu \nu_{R,i} ) \\ \nonumber
\end{eqnarray}
where
\begin{eqnarray}
\label{eqn-L-model-1-V}
V(H,\, \phi_\chi,\, \phi_s) = && m_H^2 |H|^2 + m_{\phi_\chi}^2 |\phi_\chi|^2  + m_{\phi_s}^2 |\phi_s|^2 + \lambda_H |H|^4 \nonumber \\
&& + \lambda_{\phi_\chi} |\phi_\chi|^4 + \lambda_{\phi_s} |\phi_s|^4
   + \lambda_{\chi H}  |\phi_\chi|^2  |H|^2 + \lambda_{s H}  |\phi_s|^2  |H|^2 \\ \nonumber
&& + \lambda_{\chi s}  |\phi_\chi|^2  |\phi_s|^2   + \lambda^\prime_{\chi s}  \Big( (\phi^*_\chi \phi_s)^2 + h.c. \Big)
\end{eqnarray}
with $D^\prime_{\mu} = \partial_\mu - i g_{Y^\prime} Y^\prime Z^\prime_{\mu}, \, F^\prime_{\mu\nu}=\partial_\mu Z^\prime_{\nu} - \partial_\nu Z^\prime_{\mu}$ and  $f=u,d,\ell$.
To satisfy the condition \textbf{IV-Collider}, we set $\lambda_{\chi H}=\lambda_{s H}=0$ and choose proper $m_H^2$ and $\lambda_H$ so that the properties of the $H$-dominated scalar are the same as those of the SM Higgs boson.
The term $\lambda^\prime_{\chi s}  \Big( (\phi^*_\chi \phi_s)^2 + h.c. \Big)$ in Eq.(\ref{eqn-L-model-1-V}) can generate an important mass
splitting between the real and imaginary components in $\phi_\chi$. To be more explicit, we have
\begin{eqnarray}
\phi_s \to \frac{1}{\sqrt{2}} (\phi_{R,s} + i \, \phi_{I,s}), \quad \phi_\chi \to \frac{1}{\sqrt{2}} (\phi_{R,\chi} + i \, \phi_{I,\chi}),
\end{eqnarray}
before $v_s$ is developed, and consequently
\begin{eqnarray}
( \phi^*_\chi \phi_s )^2 + h.c. &=& \frac{1}{4} \Big( (\phi_{R,\chi} - i \, \phi_{I,\chi})(\phi_{R,s} + i \, \phi_{I,s}) \Big)^2 + h.c.\\ \nonumber
&=& \frac{1}{4} \Big( (\phi_{R,\chi} \phi_{R,s}  + \phi_{I,\chi} \phi_{I,s} ) + i \, (\phi_{R,\chi} \phi_{I,s} - \phi_{I,\chi}\phi_{R,s}) \Big)^2 + h.c. \\ \nonumber
&=& \frac{1}{2} (\phi_{R,\chi} \phi_{R,s}  + \phi_{I,\chi} \phi_{I,s} )^2 - \frac{1}{2} (\phi_{R,\chi} \phi_{I,s} - \phi_{I,\chi}\phi_{R,s} )^2 \\ \nonumber
&\supset& \frac{1}{2} \phi_{R,s}^2 (\phi_{R,\chi}^2 - \phi_{I,\chi}^2).
\end{eqnarray}
When $\phi_s$ acquires a vev $v_s$ we perform the following replacement
\begin{eqnarray}
\phi_{R,s} \to v_s + \phi_{R,s},
\end{eqnarray}
which implies
\begin{eqnarray}
\lambda^\prime_{\chi s}  \Big( (\phi^*_\chi \phi_s)^2 + h.c. \Big) \supset \frac{1}{2} \, \lambda^\prime_{\chi s} v_s^2 (\phi_{R,\chi}^2 - \phi_{I,\chi}^2).
\end{eqnarray}
This induces a mass splitting $\Delta m^2_{\chi,\chi^\prime} = m^2_{\chi^\prime}-m^2_\chi = 2 \, \lambda^\prime_{\chi s} v_s^2$ with $\chi^\prime\equiv \phi_{R,\chi}$ and $\chi\equiv \phi_{I,\chi}$. Note that $\chi$ corresponds to the DM candidate for a positive $\lambda^\prime_{\chi s}$. In our discussion, we use the minimization condition of the scalar potential to replace $m_{\phi_s}$ in Eq.(\ref{eqn-L-model-1-V}) by $v_s$ as theoretical input parameter.

\begin{figure}[t]
\begin{center}
\includegraphics[width=4.4cm]{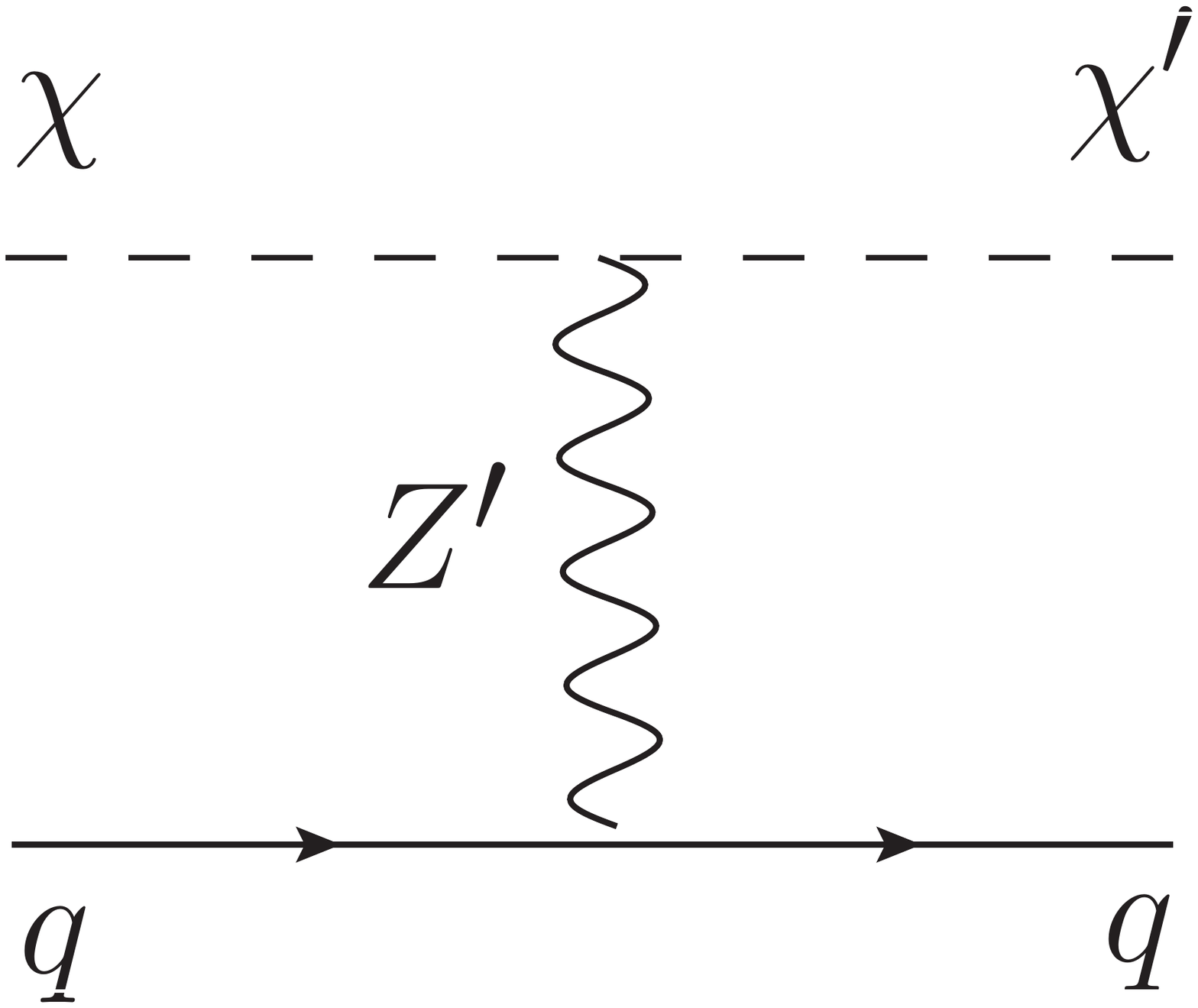}
\includegraphics[width=4.5cm]{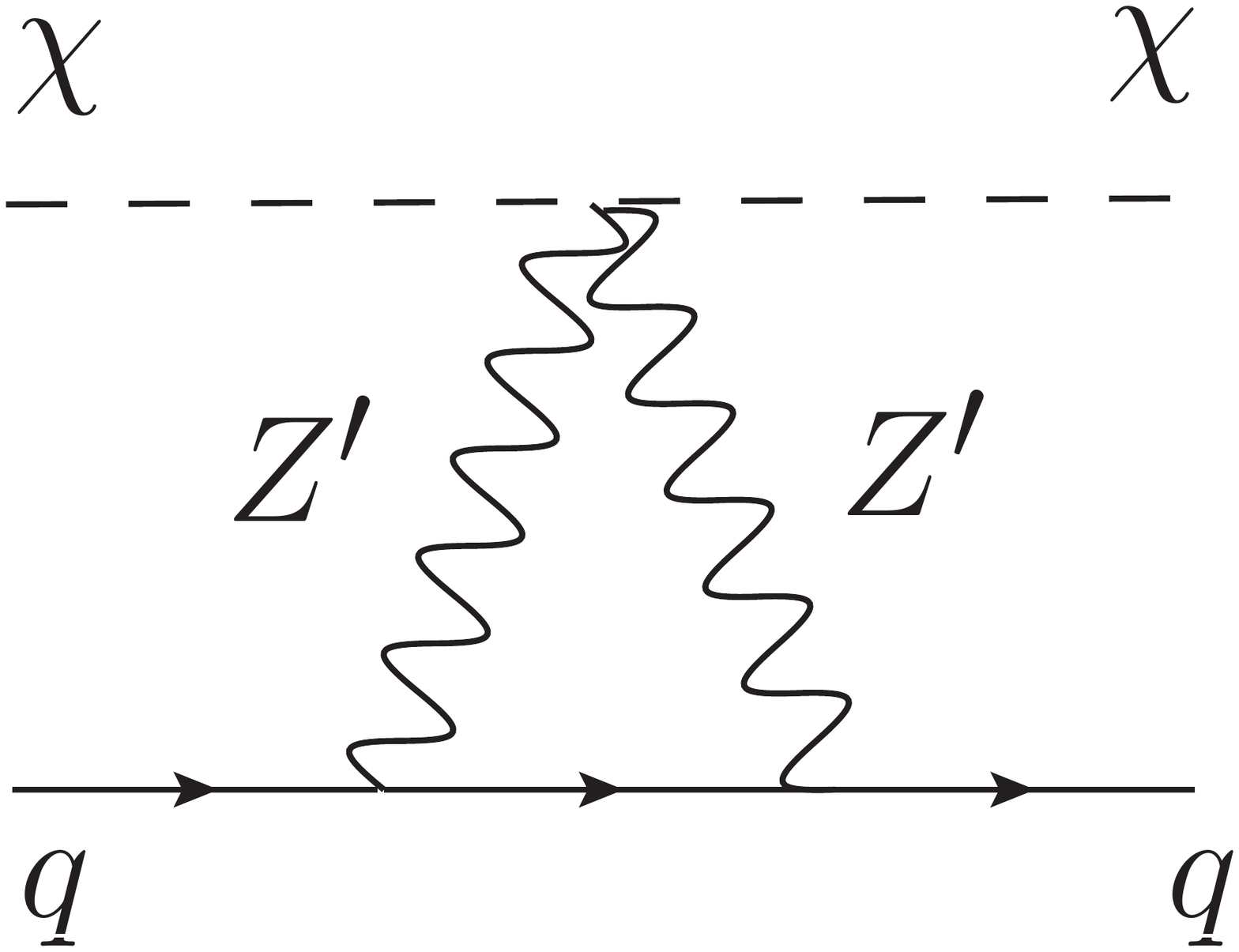}
\includegraphics[width=4.5cm]{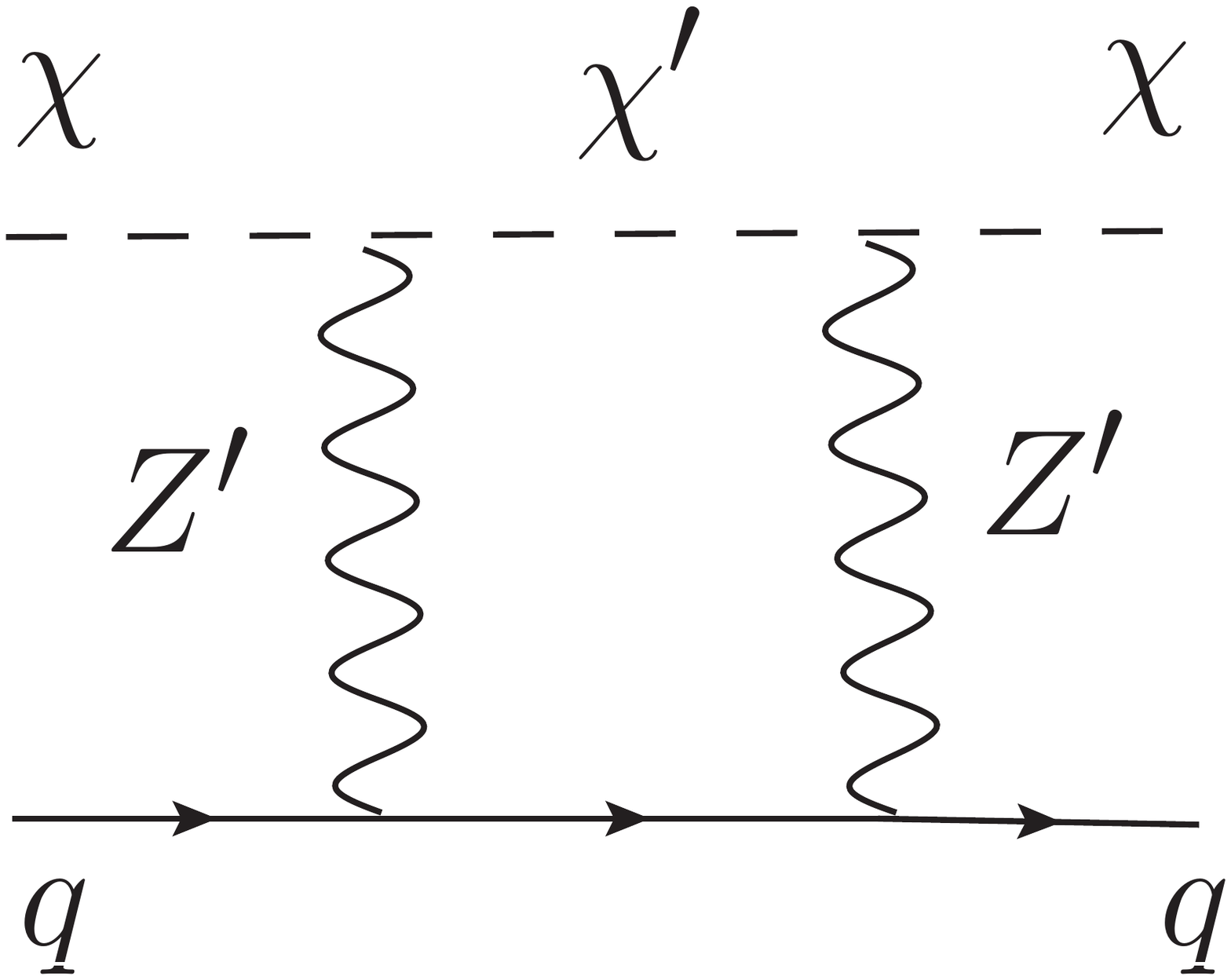}
\vspace{-1cm}
\caption{DM-nucleon scattering processes in model $G_{SM} \times U(1)_{Y^\prime}$. {\textbf{Left}}: inelastic scattering proceeds at tree level; {\textbf{Middle, Right}}: elastic scattering starts from 1-loop level.}
\label{fig-1-DD}
\end{center}
\end{figure}

The mass splitting $\Delta m_{\chi,\chi^\prime}$ has an important implication on the DM-nucleon scattering. To see this, we expand the covariant kinetic term in eq.(\ref{eqn-L-model-1}) as follows:
\begin{eqnarray}
| D^\prime_{\mu} \phi_\chi |^2 &=& (\partial^\mu \phi_\chi - i g_{Y^\prime} Y^\prime_{\phi_\chi} Z^{\prime \mu} \phi_\chi ) (\partial_\mu \phi^*_\chi + i g_{Y^\prime} Y^\prime_{\phi_\chi} Z^\prime_{\mu} \phi^*_\chi ) \\ \nonumber
&=&  \partial^\mu \phi_\chi \partial_\mu \phi^*_\chi + (g_{Y^\prime} Y^\prime_{\phi_\chi})^2 Z^{\prime \mu} Z^\prime_{\mu} |\phi_\chi|^2 + i g _{Y^\prime} Y^\prime_{\phi_\chi}  Z^\prime_{\mu} ( \phi^*_\chi \partial^\mu \phi_\chi - \phi_\chi \partial^\mu \phi^*_\chi) \\ \nonumber
&=& \partial^\mu \phi_\chi \partial_\mu \phi^*_\chi + (g_{Y^\prime} Y^\prime_{\phi_\chi})^2 Z^{\prime \mu} Z^\prime_{\mu} (\phi_{R,\chi}^2 + \phi_{I,\chi}^2 ) - 2 \, g _{Y^\prime}  Y^\prime_{\phi_\chi} Z^\prime_{\mu} (\phi_{R,\chi} \partial^\mu \phi_{I,\chi} - \phi_{I,\chi} \partial^\mu \phi_{R,\chi}).
\end{eqnarray}
where the third term indicates that there is no $Z^\prime\chi\chi$ or $Z^\prime\chi^\prime\chi^\prime$ interaction, but
only $Z^\prime \chi^\prime\chi$ interaction.
This observation implies that if the mass splitting
$\Delta m_{\chi,\chi^\prime}= m_{\chi^\prime} - m_\chi $ is much larger than
the typical DM kinetic energy
\begin{eqnarray}
E_{\chi,kin.}\sim \frac{1}{2}m_\chi v_\chi^2 \sim 0.5 \times (3\times10^3 \,{\rm GeV}) \times (10^{-3})^2 \sim 1 \,{\rm MeV},
\end{eqnarray}
the inelastic scattering process $\chi N \to \chi^\prime N$ (proceeded by the {\textbf{Left}} panel of Fig.\ref{fig-1-DD}) with
$N$ denoting nucleon is kinematically forbidden.

On the contrary, the elastic
scattering processes $\chi N \to \chi N$ start from 1-loop level, including triangle diagram involving $\chi \chi Z^\prime Z^\prime$ interaction (see {\textbf{Middle}} panel of Fig.\ref{fig-1-DD}), as well as box diagrams involving double $U(1)_{Y^\prime}$ gauge boson and light quark propagator (see {\textbf{Right}} panel of Fig.\ref{fig-1-DD}). Obviously, the rates of the elastic processes are suppressed by a loop factor, light quark mass as well as small coupling between $Z^\prime$ and light quarks. To estimate the scattering rate, we first choose the degenerate limit $\Delta m_{\chi,\chi^\prime} = 0$ in which case the scattering $\chi N \to \chi^\prime N$ proceeds mainly via the tree-level $Z^\prime$-mediated diagram. We utilize the package \textbf{micrOMEGAs} \cite{Belanger:2014vza,Belanger:2014hqa} to calculate the scattering rate and the result is about $9 \times 10^{-43} \,{\rm cm^2}$ for $m_{\chi} \simeq m_{Z^\prime} = 3 \,{\rm TeV}$ and $g_{Y^\prime} \simeq 0.2$, which is the benchmark parameter choice needed to produce the observed DM relic density while explaining the DAMPE peak (see discussion below). We conclude that, although this rate is about $40$ times larger than the current bound from the DM direct search experiments such as XENON-1T \cite{Aprile:2017iyp} and PandaX-II \cite{Cui:2017nnn}, the non-degenerate case $\Delta m_{\chi,\chi^\prime} \gg 1 \,{\rm MeV}$ can easily turn off the inelastic scattering and make our model pass the constraint \textbf{III-DD}.

\begin{table}
\caption{The framework $G_{SM} \times U(1)_{Y^\prime}$ confronting the four conditions.}

\vspace{-0.5cm}

\begin{center}
\begin{tabular}{|c|c|c|}
\hline
Condition & Result  &  Details \\
\hline \hline
\textbf{I-ID} & $\surd$ & \makecell{ $\chi\chi \to Z^\prime Z^\prime \to \ell \bar{\ell}\ell^\prime \bar{\ell^\prime}$ with $\ell,\ell^\prime=e,\mu,\tau$ \\ with $m_\chi \sim m_{Z^\prime} \sim 2 \times 1.5$ (TeV). \\ $Br(Z^\prime \to l_i^+ l_i^-) \simeq 2/9 $ and  $Br(Z^\prime \to \nu_i \bar{\nu}_i) \simeq 1/9$ for $i= e, \mu, \tau$, \\
$Br(Z^\prime\to q\bar{q})$ is small due to the small charge assignment for quarks. } \\
\hline
\textbf{II-RD} & $\surd$ & Same as \textbf{I-ID} since the annihilation is $s$-wave dominated.  \\
\hline
\textbf{III-DD} & $\surd$ & \makecell{ $\Delta m_{\chi,\chi^\prime}=m_{\chi^\prime}-m_\chi \gg E_{\chi,kin}$ forbids \\ tree-level inelastic scattering $\chi N \to \chi^\prime N$ via $t$-channel exchange of $Z^\prime$.} \\
\hline
\textbf{IV-Collider} & $\times$ & \makecell{Excluded by $q \bar{q} \to Z^\prime \to \ell \bar{\ell}$ at the LHC \cite{Aaboud:2017buh}.} \\
\hline
\end{tabular}
\end{center}
\label{table-model-1-summary}
\end{table}

\begin{figure}[t]
\begin{center}
\includegraphics[width=9.5cm]{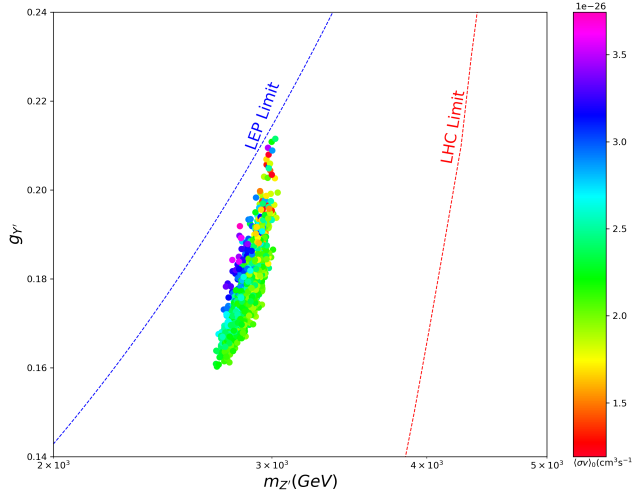}

\vspace{-0.8cm}

\caption{Samples surviving the constraints in Eq.(\ref{Constraints}), which are projected on $m_{Z^\prime}-g_{Y^\prime}$ plane.
The bounds on $m_{Z^\prime}$ from LEP experiment and LHC experiment are also shown with the left regions of the curves being excluded.}
\label{fig:fig-model-1-summary}
\end{center}
\end{figure}

For the numerical calculations, we use \textbf{SARAH} \cite{Staub:2015kfa} to implement the model, \textbf{SPheno} \cite{Porod:2003um,Porod:2011nf} to calculate the mass spectrum and \textbf{micrOMEGAs} \cite{Belanger:2014vza,Belanger:2014hqa} to calculate DM relic abundance in which the threshold effects may be important for $m_\chi \sim m_{Z^\prime}$ \cite{PhysRevD.43.3191}. We set $ \lambda_{\phi_\chi} = \lambda_{\chi s} = \lambda_{\chi H} = \lambda_{s H} = 0$, and scan by the package \textbf{Easyscan-HEP} \cite{Easyscan}, which is based on Markov Chain Monte Carlo (MCMC) sampling technique \cite{MCMC},  the following parameters space \footnote{In our calculation we have included the radiative correction to $m_\chi$, i.e. we have $m_\chi \neq m_{\phi_\chi}$ even when $\lambda_{\chi s} = \lambda'_{\chi s} =0$.}
\begin{eqnarray}
0 \leq \lambda_{\phi_s},\ g_{Y^\prime}, \,  \lambda^\prime_{\chi s} \leq 1, \  2.9 \,{\rm TeV} \leq m_{\phi_\chi} \leq 3.1 \,{\rm TeV}, \
1 \,{\rm TeV} \leq v_s \leq 5 \,{\rm TeV}.
\end{eqnarray}
We impose the constraints \cite{Ade:2015xua}
\begin{eqnarray}
m_\chi &\in& 3000 \pm 10 \, {\rm GeV}, \   m_{Z^\prime} \in (2.7 \,{\rm TeV}, 3 \,{\rm TeV}),\ m_{\phi_{R,s}},m_{\chi'} > 3 \,{\rm TeV}, \nonumber \\
\Omega_{\chi} h^2 &\in& 0.1187 \pm 0.01198, \  \langle \sigma v \rangle_0 \gtrsim 1\times 10^{-26}\, {\rm cm^3/s} \label{Constraints}
\end{eqnarray}
where both of the two physical scalars $S\equiv \phi_{R,s}$ and $\chi^\prime$ are required to
be heavier than $3 \,{\rm TeV}$ so that they do not affect significantly the explanation of the DAMPE data.
The scalar $S$ may decay into $Z^{\prime (*)} Z^{\prime (*)}$ or neutrinos via new Yukawa interactions
under $U(1)_{Y^\prime}$ such as $\phi_s \overline{\nu^c_{R,1}} \nu_{R,23}, \, \phi_s \nu^c_{R,i} \nu^c_{R,i}$ \cite{Lee:2010hf},
and the scalar $\chi^\prime$ may decay via $\chi^\prime \to \chi + Z^{\prime (*)}$ through the $U(1)_{Y^\prime}$ gauge interaction.

The model $G_{SM} \times U(1)_{Y^\prime}$ confronting the four conditions is shown in Table.\ref{table-model-1-summary}, and the numerical results are provided in Fig.\ref{fig:fig-model-1-summary} on the plane of $m_{Z^\prime}$ versus $g_{Y^\prime}$, where the left regions of LEP and LHC bounds have
been excluded \footnote{In the model $G_{SM} \times U(1)_{Y^\prime}$, the LEP bound on $m_{Z^\prime}$ comes from the process $e^+ e^- \to Z^\prime \to \mu^+\mu^-$,
which requires $m_{Z^\prime} \gtrsim g_{Y^\prime}/0.1 \times 1.4 \,{\rm TeV}$ \cite{Lee:2010hf}. The LHC bound on the mass arises from the ATLAS search for
di-lepton signal at 13-TeV LHC \cite{Aaboud:2017buh}, and we get the corresponding curve in Fig.\ref{fig:fig-model-1-summary} by simulating the analysis in \cite{Aaboud:2017buh}. For details of the simulation, see Section 10.3 and Fig.6 of \cite{Aaboud:2017buh}.}.
This figure indicates that for $m_\chi \sim m_{Z^\prime} \sim 2 \times 1.5$ (TeV), $g_{Y^\prime}$ should be around $0.2$ to produce the observed DM relic density and meanwhile to explain the DAMPE peak. In this case, $\langle \sigma v \rangle_0$ varies between $1\sim3.5 \times 10^{-26} \,{\rm cm^3/s}$.
This figure also shows that the samples with $\langle \sigma v \rangle_0 \gtrsim 1 \times 10^{-26} \,{\rm cm^3/s}$
to explain the DAMPE data can survive the LEP bound from $e^+ e^- \to Z^{\prime \ast} \to \mu^+ \mu^-$ \cite{Lee:2010hf}, but fail
to pass the LHC limits on $q \bar{q} \to Z^\prime \to \ell \bar{\ell}$ \cite{Aaboud:2017buh}. Note that the LHC exclusion of
this model originates from the chosen quark quantum numbers under $U(1)_{Y^\prime}$, which is due to the gauge anomaly free requirement and the assumption of $Z^\prime$ democratic decays into $e, \mu, \tau$.

\section{\label{Section-Model-2}$G_{SM} \times U(1)_{Y^\prime} \times U(1)_{Y^{\prime \prime}}$ framework}

To pass the LHC constraints frustrating the minimal framework $G_{SM} \times U(1)_{Y^\prime}$, we extend it with another $U(1)_{Y^{\prime \prime}}$ gauge group which mixes with the $U(1)_{Y^\prime}$ group, and with one additional complex scalar $\phi_d$ to break the $U(1)_{Y^{\prime \prime}}$ symmetry. The particle contents in this model with chiral anomaly cancellation are shown in Table.\ref{table-model-2}. Note that $\phi_\chi$ is no longer charged under the previous $U(1)_{Y^\prime}$ and only $\phi_\chi,\phi_d$ are charged under $U(1)_{Y^{\prime \prime}}$.

\begin{table}
\caption{Particle contents in $G_{SM} \times U(1)_{Y^\prime} \times U(1)_{Y^{\prime \prime}}$ model.}
\begin{center}
\begin{tabular}{|c|c|c|c|c|c|c|c|}
\hline
Name 				& Spin	&  Gen.	 & $SU(3)_C$  & $SU(2)_L$ & $U(1)_Y$   &  $U(1)_{Y^\prime}$  & $U(1)_{Y^{\prime \prime}}$  \\
\hline \hline
$H$						& 0 		& 1 	& {\bf 1} 					& {\bf 2} 		& -$\frac{1}{2}$ 	&  0 &  0 \\
\hline
$Q$						& 1/2 		& 3 	& {\bf 3} 					& {\bf 2} 		&  $\frac{1}{6}$ 	&  $\frac{1}{3}$ &  0 \\
$d_R^*$ 				& 1/2 		& 3 	& {\bf $\bar{\bf 3}$}	& {\bf 1} 		&  $\frac{1}{3}$ 	& -$\frac{1}{3}$ &  0 \\
$u_R^*$					& 1/2 		& 3 	& {\bf $\bar{\bf 3}$}	& {\bf 1} 		& -$\frac{2}{3}$	& -$\frac{1}{3}$ &  0 \\
\hline
$L_1$					& 1/2 		& 1 	& {\bf 1}					& {\bf 2} 		& -$\frac{1}{2}$ 	&  3 &  0 \\
$L_{\{2,3\}}$			& 1/2 		& 2 	& {\bf 1} 					& {\bf 2} 		& -$\frac{1}{2}$ 	& -3 &  0 \\
$\ell^*_{R,1}$			& 1/2 		& 1 	& {\bf 1} 					& {\bf 1} 		& 1 					& -3 &  0 \\
$\ell^*_{R,\{2,3\}}$	& 1/2 		& 2 	& {\bf 1}					& {\bf 1} 		& 1 					& 3 &  0 \\
\hline \hline	
$\nu^*_{R,1}$			& 1/2 		& 1	& {\bf 1}					& {\bf 1} 		& 0 					& -3 &  0 \\
$\nu^*_{R,\{2,3\}}$	& 1/2 		& 2 	& {\bf 1} 					& {\bf 1} 		& 0 					&  3 &  0 \\
$\phi_s$ 				& 0		& 1 	& {\bf 1} 					& {\bf 1} 		& 0 					&  6 &  0 \\
\hline
$\phi_\chi$				& 0		& 1 	& {\bf 1} 					& {\bf 1} 		& 0 					&  0 &  $Y^{\prime \prime}_{\phi_\chi}$ \\
$\phi_d$					& 0		& 1 	& {\bf 1} 					& {\bf 1} 		& 0 					&  0 &  $Y^{\prime \prime}_{\chi_d}$     \\
\hline
\end{tabular}
\end{center}
\label{table-model-2}
\end{table}

The most relevant Lagrangian to explain the DAMPE data include
\begin{eqnarray}
\label{eqn-L-model-2}
{\mathcal L} \supset && | D_{\mu} \phi_\chi |^2 + | D_{\mu} \phi_s |^2 + | D_{\mu} \phi_d |^2 - V(H,\, \phi_\chi,\, \phi_s, \, \phi_d) \\ \nonumber
&& - \frac{1}{4} |F^\prime_{\mu\nu}|^2 - \frac{1}{4} |F^{\prime \prime}_{\mu\nu}|^2 - \frac{\kappa}{2} F^{\prime \mu\nu} F^{\prime \prime}_{\mu\nu}
\end{eqnarray}
where \footnote{For a $U(1)_1 \times U(1)_2$ gauge theory, $D_\mu$ is usually presented in a compact form \cite{delAguila:1988jz,Luo:2002iq,Chankowski:2006jk}, which is given by $ D_\mu \phi = \left [ \partial_\mu + i \sum_a \sum_b Y^a_\phi g_{ab} B_\mu^b \right ] \phi$ with $a,b=1,2$ and $g$ corresponding to a coupling matrix.
In our formulae, we expand the summation explicitly. }
\begin{eqnarray}
\label{eqn-L-model-2-V}
D_{\mu} \phi = && \left [ \partial_\mu + i Y_{\phi}^\prime ( g_{Y^\prime Y^\prime} B_\mu^\prime + g_{Y^\prime Y^{\prime \prime}} B_\mu^{\prime \prime} ) +
i Y_{\phi}^{\prime \prime} ( g_{Y^{\prime \prime} Y^\prime} B_\mu^\prime + g_{Y^{\prime \prime} Y^{\prime \prime}} B_\mu^{\prime \prime} ) \right ] \phi, \nonumber \\
F^\prime_{\mu\nu}= && \partial_\mu B^\prime_{\nu} - \partial_\nu B^\prime_{\mu}, \quad F^{\prime \prime}_{\mu\nu}=\partial_\mu B^{\prime \prime}_{\nu} - \partial_\nu B^{\prime \prime}_{\mu}, \nonumber \\
V(H,\, \phi_\chi,\, \phi_s, \, \phi_d) = && m_H^2 |H|^2 + m_{\phi_\chi}^2 |\phi_\chi|^2  + m_{\phi_s}^2 |\phi_s|^2 + m_{\phi_d}^2 |\phi_d|^2 \\ \nonumber
&& + \lambda_H |H|^4 + \lambda_{\phi_\chi} |\phi_\chi|^4 + \lambda_{\phi_s} |\phi_s|^4 + \lambda_{\phi_d} |\phi_d|^4 \\ \nonumber
&& + \lambda_{\chi H}  |\phi_\chi|^2  |H|^2 + \lambda_{s H}  |\phi_s|^2  |H|^2 + \lambda_{d H}  |\phi_d|^2  |H|^2 \\ \nonumber
&&  + \lambda_{\chi s}  |\phi_\chi|^2  |\phi_s|^2   + \lambda_{\chi d}  |\phi_\chi|^2  |\phi_d|^2 + \lambda_{s d}  |\phi_s|^2  |\phi_d|^2. \nonumber
\end{eqnarray}
In the above expressions, $Y_\phi^\prime,Y_\phi^{\prime \prime}$ denote the $U(1)_{Y^\prime}$ and $U(1)_{Y^{\prime\prime}}$ charge of $\phi$, respectively, and $g_{ab}$ with $a,b = Y^\prime, Y^{\prime \prime}$ are the coupling constants, while $\kappa$
parameterizes the kinetic mixing between $U(1)_{Y^\prime}$ and $U(1)_{Y^{\prime \prime}}$. Note that the appearance a coupling matrix $g$ instead 
of one single gauge coupling for each U(1) group is a special feature of the theory with multiple U(1)¡¯s  \cite{delAguila:1988jz,Luo:2002iq,Chankowski:2006jk}.
In fact, even if $g$ is not included in the tree-level Lagrangian, the kinematic mixing term $\kappa$ and the
$g_{ab}$ couplings will be generated in the effective action through radiative corrections if there exist fields charged under both the groups. 
Also note that we can choose
$Y^{\prime \prime}_{\phi_\chi} \neq Y^{\prime \prime}_{\phi_d}$ to forbid $\lambda^\prime_{\chi s}$ terms in Eq.(\ref{eqn-L-model-1-V}),
but for comparison we take $Y^{\prime \prime}_{\phi_\chi} = Y^{\prime \prime}_{\phi_d} = 6$ as we did for previous model and set
$\lambda^\prime_{\chi d}=0$ since we no longer need to rely on non-zero $\Delta m_{\chi,\chi^\prime}$ to forbid tree-level $Z^\prime$-mediated DM-nucleon scattering in escaping the DM direct search constraints (see discussion below).

In the following we denote $g_{Y^\prime}\equiv g_{Y^\prime Y^\prime}, g_{Y^{\prime \prime}} \equiv g_{Y^{\prime \prime} Y^{\prime \prime}}$,
$\varepsilon \equiv g_{Y^{\prime} Y^{\prime \prime}} = g_{Y^{\prime \prime} Y^{\prime}}$, and neglect the explicit kinematic mixing term $\kappa$ for simplicity \footnote{As illustrated
in \cite{Chankowski:2006jk,Feldman:2007wj}, including the mixing corresponds to a redefinition of the fields and
does not essentially change the results in this work.}.  Then in the basis $(B^\prime, B^{\prime \prime})$, the mass matrix of the new
gauge bosons is given by
\begin{eqnarray}
M_V^2 = 36 \times \left(
\begin{array}{cc}
  g_{Y^\prime}^2 v_s^2 + \varepsilon^2 v_d^2 \quad & \quad \varepsilon  (g_{Y^\prime} v_s^2 + g_{Y^{\prime \prime}} v_d^2) \quad \\
  \varepsilon (g_{Y^\prime} v_s^2 + g_{Y^{\prime \prime}} v_d^2) \quad & \quad \varepsilon^2 v_s^2 + g_{Y^{\prime \prime}}^2 v_d^2 \quad \\
\end{array}
\right).
\end{eqnarray}
After diagonalization of the matrix one can obtain the mass eigenstates $Z^\prime$ and $Z^{\prime \prime}$, which are
$B^\prime_\mu$ and $B^{\prime \prime}_\mu$ dominated, respectively.
Similar to the minimal framework of $G_{SM} \times U(1)_{Y^\prime}$, charge assignments in Table \ref{table-model-2} make $Z^\prime$ and $Z^{\prime \prime}$ decay dominantly into $l_i^+ l_i^-$ and $\nu_i \bar{\nu}_i$ if the $\phi_{s,d}$ dominated CP-even scalars are sufficiently heavy. In this new framework, we will focus on the complex scalar DM $\chi\equiv \phi_{\chi}$ and utilize the process $\chi\chi \to Z^{\prime \prime}Z^{\prime \prime} \to \ell \bar{\ell}\ell^\prime \bar{\ell^\prime}$ with $\ell,\ell^\prime=e,\mu,\tau$ to explain the DAMPE peak. The sensitive parameters, as we discussed in the minimal framework, are $m_{\chi}$, $m_{Z^{\prime \prime}}$, $g_{Y^{\prime \prime}}$ and also $ \varepsilon$. The parameter $\varepsilon$ affects not only the masses and the $B^\prime_\mu, B^{\prime \prime}_\mu$ components of the new gauge bosons, but also the $\chi N \to \chi N$ scattering rate which proceeds via the tree level $t$-channel exchange of $Z^\prime$ or $Z^{\prime \prime}$.

 Like what we did in last section, we explore the parameter space of the non-minimal framework $G_{SM} \times U(1)_{Y^\prime} \times 
 U(1)_{Y^{\prime \prime}}$ to interpret the DAMPE peak. Firstly we set
$\lambda_{\{\chi,s,d\},H} = \lambda_{\phi_\chi} = \lambda_{\chi s} = \lambda_{\chi d} = \lambda_{s d} = 0$, $g_{Y^\prime} = 0.1$ to simplify our analysis
and also to ensure that the $H$-dominated CP-even scalar acts as the SM Higgs boson.
Then we scan the following parameter space
\begin{eqnarray}
0 \leq \lambda_{\phi_s},\, \lambda_{\phi_d},\, g_{Y^{\prime \prime}} , \,\varepsilon \leq 1, \quad
2.9 \,{\rm TeV} \leq m_{\phi_\chi} \leq 3.1 \,{\rm TeV}, \quad  6 (2)\,{\rm TeV} \leq v_s (v_d) \leq 10 (5) \,{\rm TeV}, \nonumber 
\end{eqnarray}
with the following requirements
\begin{eqnarray}
&& m_\chi  \in 3000 \pm 10 \, {\rm GeV}, \quad m_{Z^{\prime \prime}}  \in (2.7 \,{\rm TeV}, 3 \,{\rm TeV}),\quad  m_{\phi_{R,s}}, m_{\phi_{R,d}} > 3 \,{\rm TeV}, \nonumber \\
&& m_{Z^\prime} > 4 \,{\rm TeV}, \quad \Omega_{\chi} h^2 \in 0.1187 \pm 0.01198, \quad  \langle \sigma v \rangle_0 \gtrsim 1\times 10^{-26}\, {\rm cm^3/s}.
\end{eqnarray}

\begin{figure}[t]
\begin{center}
\includegraphics[width=9.5cm]{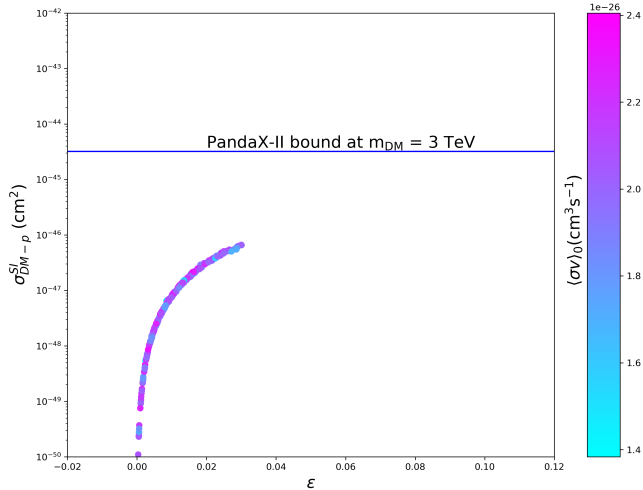}
\caption{DM-nucleon scattering rate as a function of $varepsilon$ in the framework $G_{SM} \times U(1)_{Y^\prime} \times U(1)_{Y^{\prime \prime}}$ with the color bar denoting DM annihilation rate at present time.}
\label{fig:fig-model-2-summary}
\end{center}
\end{figure}

For the surviving samples, we present the details of how the four conditions \textbf{I-ID, II-RD, III-DD, IV-Collider} are satisfied:
\begin{itemize}
\item \textbf{I-ID}:  $g_{Y^{\prime \prime}}\sim 0.2$ with $m_\chi \sim m_{Z^{\prime \prime}}\sim 3.0$ TeV can generate $\langle \sigma v \rangle_0 \gtrsim 1 \times 10^{-26} \,{\rm cm^3/s}$, followed by $Z^{\prime \prime}$ decaying into nearly equal $e,\mu,\tau$ Brs. In this case, we checked that for $\epsilon = 0.04$ and $m_{Z^\prime} = 5 \,{\rm TeV}$, the width of $Z^{\prime \prime}$ is about 3 GeV, which means that the lifetime of $Z^{\prime \prime}$ is short enough and does not affect the evolution of the  early Universe, and its coupling strength with leptons is about $0.1$.
\item \textbf{II-RD}: Same as \textbf{I-ID} since the DM annihilation is a $s$-wave process.
\item \textbf{III-DD}: $g_{Y^\prime}=0.1,\, \varepsilon \lesssim 0.05,\, m_{Z^\prime}\gtrsim 4\, {\rm TeV}$ can suppress the DM-nucleon scattering
 $\chi N \to \chi N$, which proceeds via the $t$-channel exchange of $Z^\prime$ or $Z^{\prime \prime}$. In Fig.\ref{fig:fig-model-2-summary},
  we project the surviving samples on the plane of $\varepsilon$ versus spin independent (SI) DM-nucleon scattering rate $\sigma_{DM-p}^{SI}$. This figure indicates
  that for $\varepsilon \lesssim 0.05$, the scattering rate is far below its current bound from PandaX-II experiment. We checked that a larger $\varepsilon$ can change significantly $m_{\chi}$ by radiative correction so that it violates the requirement of $ m_{\chi} \simeq 3 \,{\rm TeV}$.
\item \textbf{IV-Collider}: For our choice $g_{Y^\prime}=0.1, \, m_{Z^\prime}\gtrsim 4\, {\rm TeV}$ is sufficiently heavy to suppress the process
$e^+ e^- \to Z^{\prime \ast} \to \mu^+ \mu^-$ at LEP and $ p p \to Z^{\prime (\ast)} \to l_i^+ \l_i^- $ at 13-TeV LHC so that the collider constraints can be evaded (see details of collider constraints in previous section).
\end{itemize}

Before we end this section, let us emphasize that, as was pointed out in \cite{Fan2017, Liu:2017rgs}, the DM explanation of the DAMPE peak is consistent with the other DM indirect detection constraints, such as the H.E.S.S. data on the annihilation $\chi \chi \to V V \to 4 e$ \cite{Abdallah:2016ygi,Profumo:2017obk}, the Fermi-LAT data in the direction of the dwarf spheroidal galaxies \cite{Ackermann:2015zua}, the Planck CMB data which is sensitive to
energy injection to the CMB from DM annihilations \cite{Slatyer:2015jla,Slatyer:2015kla}, as well as the IceCube data on DM annihilation into neutrinos \cite{Aartsen:2017ulx}. It also survive the upper bounds of XENON-10 and XENON-100 experiments on the DM scattering
with electron \cite{Essig:2017kqs}.

\section{\label{Spectrum} $e^+e^-$ spectrum in our explanation}

\begin{figure}[t]
\begin{center}
\includegraphics[height=7cm,width=7cm]{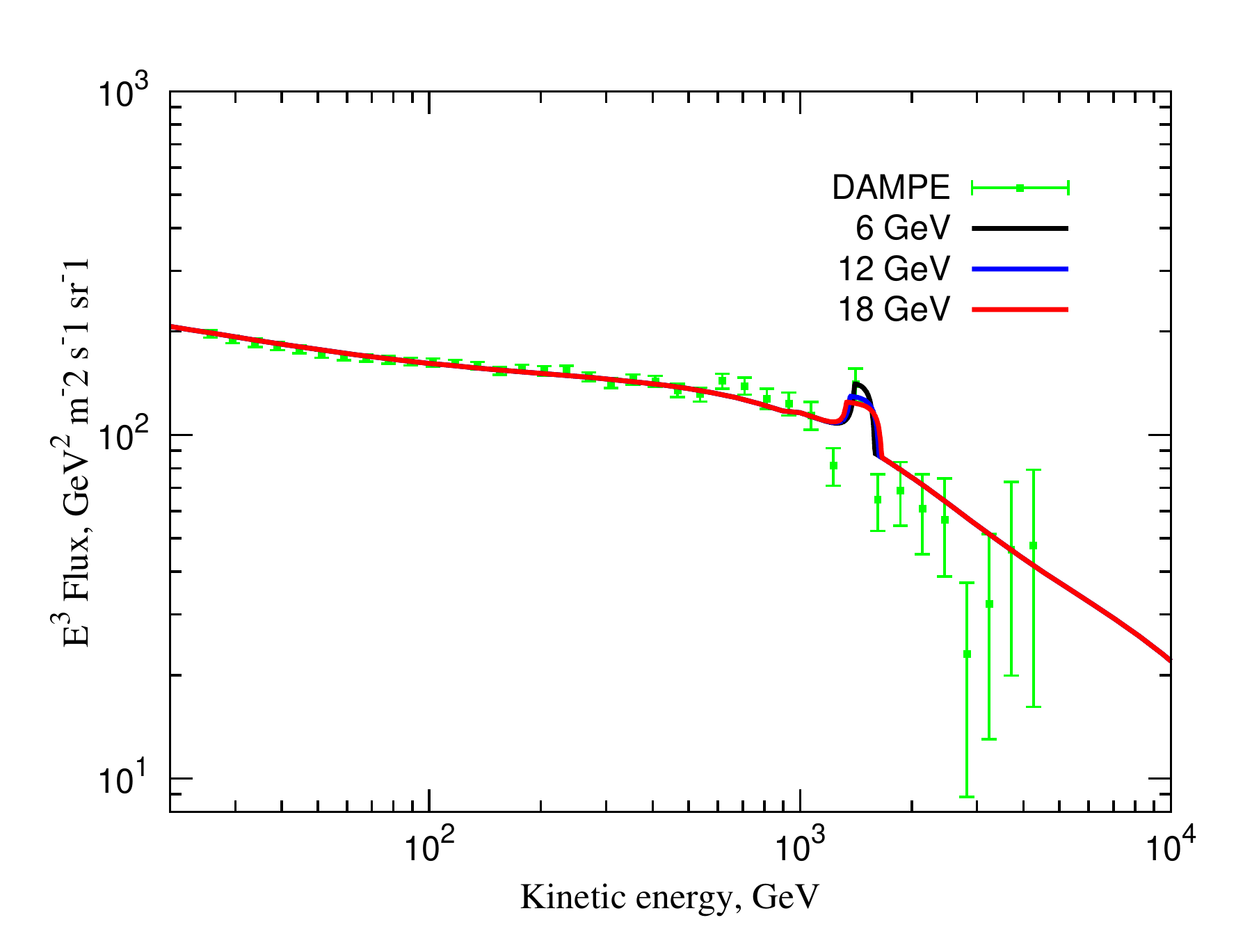}
\includegraphics[height=7cm,width=7cm]{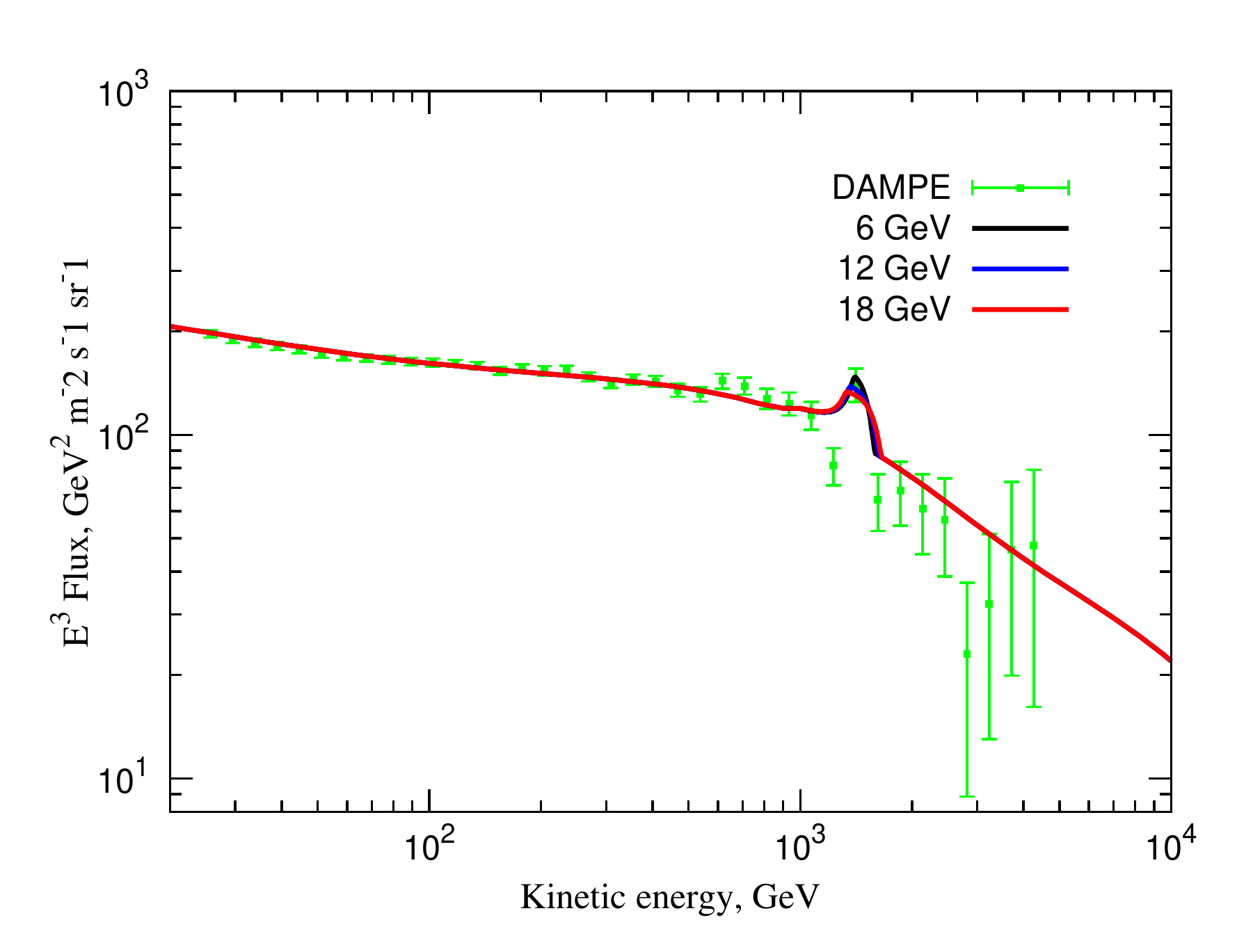}
\caption{Cosmic $e^+e^-$ spectrum generated by $\Delta m \equiv m_\chi - m_{Z'} = 6, 12, 18$ GeV, respectively, for the process $\chi \chi \to Z^{\prime \prime} Z^{\prime \prime}$ with  $Z^{\prime \prime}$ democratic decays into  $e^+e^-, \mu^+ \mu^-, \tau^+ \tau^-$ final states and $m_\chi = 3 \,{\rm TeV}$.   
The DAMPE data is also plotted for comparison. The \textbf{Left} panel corresponds to the choice $\langle \sigma v \rangle_0 = 3.2 \times 10^{-26} \,{\rm cm^3 s^{-1}}$, $M_{Halo} = \rm 1.9 \times 10^{7} \,{\rm m_\odot}$ (the subhalo mass) and $l=0.1$ kpc (the distance of the subhalo away from our solar system), while the \textbf{Right} panel corresponds to $\langle \sigma v \rangle_0 = 1.2 \times 10^{-26} \,{\rm cm^3 s^{-1}}$, $M_{Halo} = \rm 3.0 \times 10^{8} \,{\rm m_\odot}$ and $l=0.2$ kpc. }
\label{fig:fig-model-1-spectrum}
\end{center}
\end{figure}

It is well known that if DM annihilates into intermediate particles $\phi$ which subsequently decay into $e^+e^-$, the $e^+e^-$ spectrum will exhibit a box-shaped feature which is distinguishable from other astrophysical process. In order to explain the DAMPE peak by the process $\chi\chi \to Z^{\prime \prime}Z^{\prime \prime} \to \ell \bar{\ell}\ell^\prime \bar{\ell^\prime}$ considered in this work, the mass splitting $\Delta m \equiv m_\chi - m_{Z^{\prime \prime}}$ should be small (see e.g. \cite{Zu:2017dzm}) and the sub-halo where DM annihilates should locate near the Earth.

In this section, we study the $e^+e^-$ spectrum predicted by $\chi\chi \to Z^{\prime \prime}Z^{\prime \prime} \to \ell \bar{\ell}\ell^\prime \bar{\ell^\prime}$
and investigate the upper bound on $\Delta m$ in explaining the peak. We adopt the parameterization of the cosmic ray background by the formulae presented in \cite{Zu:2017dzm}
and use the \textbf{LikeDM} package \cite{Huang:2016pxg} to calculate the propagation of the $e^+e^-$ in the background. The relevant data to determine the
parameters for the background are those of the AMS-02 $e^+$ fraction and $e^+e^- $ fluxes and the DAMPE $e^+e^-$ flux.
Then we add the contribution of local sub-halo directly assuming that such component only affects the energy bin
$\rm \sim 1.5 \,{\rm TeV}$. We adopt a Navarro-Frenk-White profile \cite{Navarro:1996gj} with a truncation at the
tidal radius \cite{Springel:2008cc} for DM density distribution inside the subhalo,
and use the subhalo mass as an input to determine the profile (for the relevant method, see the appendix of \cite{Yuan2017}).
The propagation of the nearby $e^+e^-$ can be calculated analytically under the assumption of
spherically symmetric geometry and infinite boundary conditions \cite{Aharonian:1995zz}. We note that the method to obtain the DM spectrum has been described
in detail in \cite{Liu:2017rgs,Chao:2017yjg,Cao:2017sju}.

In the \textbf{Left} panel of Fig.\ref{fig:fig-model-1-spectrum} we present the spectrum of $e^+e^-$ with $\Delta m=$ 6 GeV, 12 GeV and 18 GeV, respectively.
In producing this plot, we set $m_{\chi} = 3 \,{\rm TeV}, \langle \sigma v \rangle_0 = 3.2 \times 10^{-26} \,{\rm cm^3 s^{-1}}$ and take the distance and the mass of the hypothesized sub-halo as $l = 0.1 \,{\rm kpc}, M_{Halo}= 1.9\times 10^{7} \,{\rm m_\odot}$. The parameters for the background are chosen to be the same as those in Table I of \cite{Zu:2017dzm}.
The \textbf{Right} panel of Fig.\ref{fig:fig-model-1-spectrum} is same as the \textbf{Left} panel except that we take $\langle \sigma v \rangle_0 = 1.2 \times 10^{-26} \,{\rm cm^3 s^{-1}}$, $l = 0.2 \,{\rm kpc}$ and $M_{Halo}= 3.0 \times 10^{9} \,{\rm m_\odot}$. Note that the height of spectrum is determined by  $\langle \sigma v \rangle_0$ and $M_{Halo}$, while the increase of the distance $l$ and/or $\Delta m$ will make the spectrum more box-shaped. Moreover, we note from our calculation that if the mediator decays exclusively into $e^+ e^-$, the peak is sharper than the mixed channel case considered in Fig.\ref{fig:fig-model-1-spectrum}. This is understandable since the later decays of $\mu$ and $\tau$ from the $Z^{\prime \prime}$ tend to flatten the peak. Fig.\ref{fig:fig-model-1-spectrum} shows that the process we considered can generate the peak structure to explain the DAMPE excess.

\begin{figure}[t]
\begin{center}
\includegraphics[height=7cm,width=7cm]{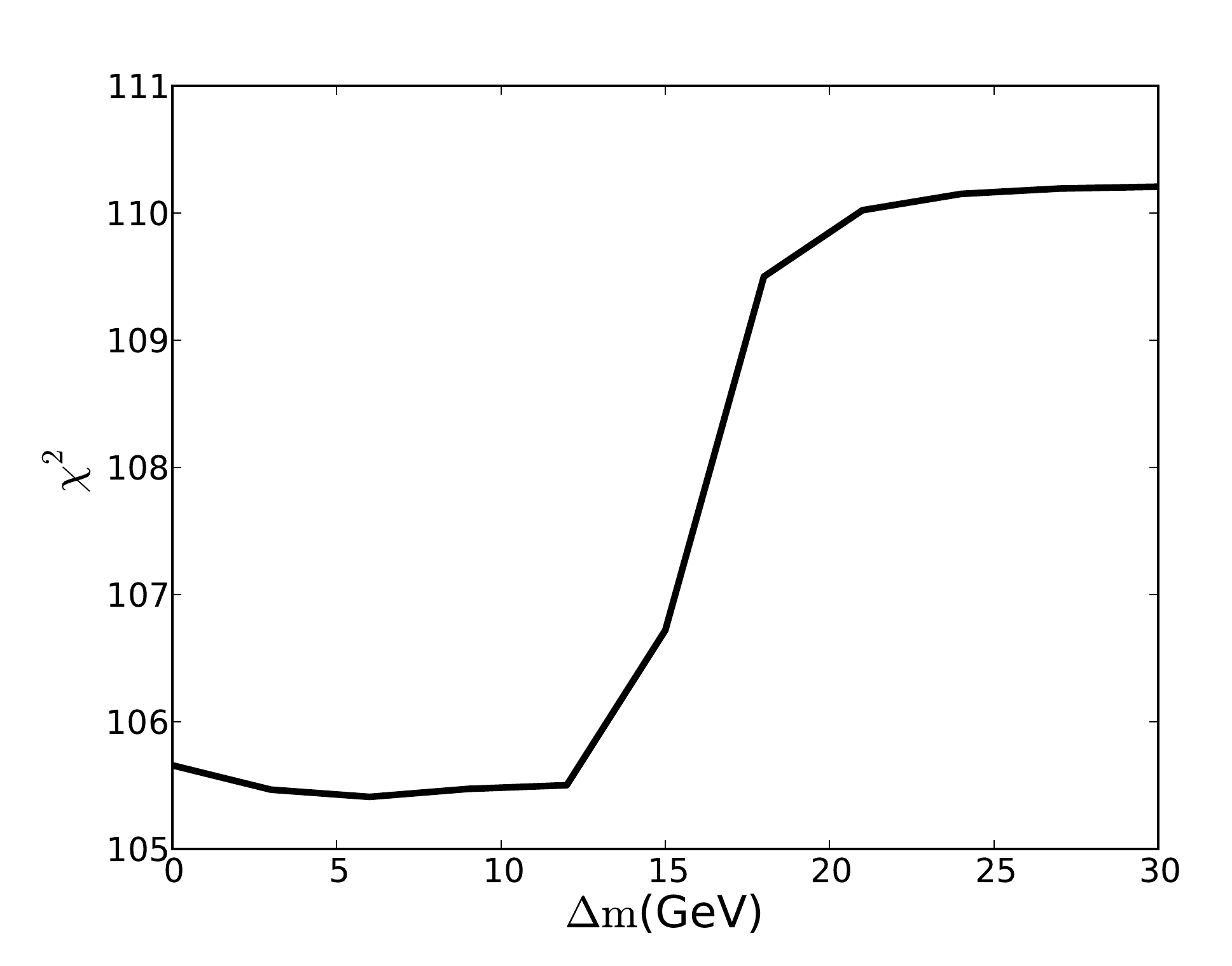}
\caption{Best fits of the $e^+e^-$ spectrum generated by the process $\chi\chi \to Z^{\prime \prime}Z^{\prime \prime} \to \ell \bar{\ell}\ell^\prime \bar{\ell^\prime}$ (with $\ell,\ell^\prime=e,\mu,\tau$) to the AMS-02 and DAMPE data, which are projected on $\Delta m -\chi^2$ plane for fixed $m_{\chi} = 3 \,{\rm TeV}$, $l = 0.1 \,{\rm kpc}$ and $M_{Halo}= 1.9 \times 10^{7} \,{\rm m_\odot}$. From this figure, one can infer that the $95\%$ upper limit on $\Delta m$ in explaining the DAMPE peak is about $17 \,{\rm GeV}$.}
\label{GF-data}
\end{center}
\end{figure}

In order to obtain the upper limit on $\Delta m$, we perform best fits of the $e^+e^-$ spectrum generated by our explanation to the AMS-02 and DAMPE data. For each chosen $\Delta m$, we fix $m_{\chi} = 3 \,{\rm TeV}, l = 0.1 \,{\rm kpc}, M_{Halo}= 1.9 \times 10^{7} \,{\rm m_\odot}$ and vary the background parameters and $\langle \sigma v \rangle_0$ to get the maximum value of the constructed likelihood function. This procedure is same as that in \cite{Zu:2017dzm}. The $\chi^2$s obtained in this way are projected on the $\Delta m -\chi^2$ plane in Fig.\ref{GF-data}. This figure indicates that the minimum value of $\chi^2$ locates at $\Delta m \simeq 5 \,{\rm GeV}$, and by requiring $(\chi^2 - \chi^2_{min} ) \leq 2.71 $ which can be used to set a $95\%$ upper bound on $\Delta m$, we get $\Delta \lesssim 17 \,{\rm GeV}$ in order to produce an acceptable peak-like $e^+e^-$ spectrum.

\section{\label{Section-Conclusion}Conclusion}

In this work we utilized the dark matter annihilation scenario to explain the tentative peak structure at around 1.4 TeV in the recently released DAMPE measurement of the total cosmic $e^+e^-$ flux between 25 GeV and 4.6 TeV. We extended $G_{SM}\equiv SU(3)_C \times SU(2)_L \times U(1)_Y$ with additional $U(1)$ gauge symmetries while keeping anomaly free to generate $\chi \chi \to Z^\prime Z^\prime \to \ell\bar{\ell}\ell^\prime\overline{\ell^\prime}$, where $\chi, Z^\prime, \ell^{(^\prime)}$ denote the scalar DM, the new gauge boson and $\ell^{(^\prime)}=e,\mu,\tau$, respectively, with $m_\chi \sim m_{Z^\prime} \sim 2 \times 1.5$ (TeV). We first illustrate that the minimal framework $G_{SM} \times U(1)_{Y^\prime}$ with the above mass choices can explain the DAMPE excess, but has been excluded by the LHC constraints from the $Z^\prime$ searches. Then we studied the non-minimal framework $G_{SM} \times U(1)_{Y^\prime} \times U(1)_{Y^{\prime \prime}}$ in which the $U(1)_{Y^{\prime \prime}}$ group mixes with the $U(1)_{Y^\prime}$ group. We showed that it can interpret the DAMPE data while passing other constraints including DM relic abundance, DM direct detection and collider bounds. We also studied the $e^+e^-$ spectrum predicted by the DM annihilation processes we consider. By performing spectrum fits to experimental data, we found that the mass splitting $\Delta m \equiv m_\chi - m_{Z^{\prime \prime}}$ between the scalar DM and the mediator vector boson should be less than about $17 \,{\rm GeV}$ to explain the observed peak.

\section*{Acknowledgement}

This work is supported by the National Natural Science Foundation of China (NNSFC) under grant No. 11575053.
This work is also supported by the Natural Science Foundation of China under grant numbers 11675147 and by the Innovation Talent project of Henan Province under grant number 15HASTIT017.


\bibliographystyle{JHEP}
\bibliography{DAMPE}

\end{document}